\documentclass[twocolumn,
secnumarabic,amssymb, nobibnotes, aps, prd, 10pt,
]{revtex4-1}
\usepackage{graphicx}
\usepackage{lineno}
\usepackage{amsmath}
\usepackage{braket}
\usepackage{color}

\begin{document}
\begin{abstract}
We report on the first detailed study of motional heating in a cryogenic Penning trap using a single antiproton. Employing the continuous Stern-Gerlach effect we observe cyclotron quantum transition rates of 6(1) quanta/h and an electric field noise spectral density below $7.5(3.4)\times 10^{-20}\,\text{V}^{2}\text{m}^{-2} \text{Hz}^{-1}$, which corresponds to a scaled noise spectral density below $8.8(4.0)\times 10^{-12}\,\text{V}^{2}\text{m}^{-2}$, results which are more than two orders of magnitude smaller than those reported by other ion trap experiments.
\end{abstract}
	
\title{Measurement of ultra-low heating rates of a single antiproton in a cryogenic Penning trap}%
\author{M. J. Borchert$^{1,2}$}\email[]{matthias.joachim.borchert@cern.ch}\
\author{P. E. Blessing$^{1,3}$} 
\author{J. A. Devlin$^{1}$} 
\author{J. A. Harrington$^{1,4}$} 
\author{T. Higuchi$^{1,5}$}
\author{J.~Morgner$^{1,2}$}
\author{C.~Smorra$^{1}$}
\author{E.~Wursten$^{1,7}$}
\author{M.~Bohman$^{1,4}$}
\author{M.~Wiesinger$^{1,4}$}
\author{A.~Mooser$^{1}$}
\author{K.~Blaum$^{4}$}
\author{Y.~Matsuda$^{5}$}
\author{C.~Ospelkaus$^{2,8}$}
\author{W.~Quint$^{3,9}$}
\author{J.~Walz$^{6,10}$}
\author{Y.~Yamazaki$^{11}$}
\author{S.~Ulmer$^1$}

\affiliation{$^1$RIKEN, Ulmer Fundamental Symmetries Laboratory, Wako, Saitama 351-0198, Japan}
\affiliation{$^2$Institut f\"ur Quantenoptik, Leibniz Universit\"at Hannover, 30167 Hannover, Germany}
\affiliation{$^3$GSI-Helmholtzzentrum f\"ur Schwerionenforschung, 64291 Darmstadt, Germany}
\affiliation{$^4$Max-Planck-Institut f\"ur Kernphysik, 69117 Heidelberg, Germany}
\affiliation{$^5$Graduate School of Arts and Sciences, University of Tokyo, Tokyo 153-8902, Japan}
\affiliation{$^6$Institut f\"ur Physik, Johannes Gutenberg-Universit\"at Mainz, 55099 Mainz, Germany}
\affiliation{$^7$CERN, 1211 Geneva 23, Switzerland}
\affiliation{$^8$Physikalisch-Technische Bundesanstalt, 38116 Braunschweig, Germany}
\affiliation{$^{9}$Ruprecht-Karls-Universit\"at Heidelberg, 69047 Heidelberg, Germany}
\affiliation{$^{10}$Helmholtz-Institut Mainz, 55099 Mainz, Germany}
\affiliation{$^{11}$Atomic Physics Laboratory, RIKEN, Wako, Saitama 351-0198, Japan}
\date{\today}
\maketitle
Quantum control techniques applied to trapped charged particles, well-isolated from environmental influences, have very versatile applications in metrology and quantum information processing. For example, elegant experiments on co-trapped laser cooled ions in Paul traps have provided highly precise state-of-the-art quantum logic clocks \cite{schmidt2005}, enabled the development of exquisite atomic precision sensors \cite{kitching2011} and the implementation of quantum information algorithms applied with highly entangled ion-crystals \cite{blatt2008}. Decoherence effects from noise driven quantum transitions, commonly referred to as \textit{anomalous heating} \cite{turchette2000,brownnutt2015}, affect the scalability of multi-ion systems, which would enable even more powerful algorithms. 
Trapped particles are also highly sensitive probes to test fundamental symmetries, and to search for physics beyond the standard model \cite{kosteleckyPenning,safronova2018}. The most precise values of the mass of the electron \cite{sturm2014} and the most stringent tests of bound-state quantum electrodynamics \cite{sturm2011} are based on precise frequency measurements on highly-charged ions in Penning traps. Measurements of the properties of trapped electrons \cite{hanneke2008} and positrons \cite{vandyck1987} provide the most sensitive tests of quantum electrodynamics and of the fundamental charge-parity-time (CPT) invariance in the lepton sector \cite{cpt,dehmelt1999}.\\ 
Our experiments \cite{smorra2015epj} make high-precision comparisons of the fundamental properties of protons and antiprotons, and provide stringent tests of CPT invariance in the baryon sector. We recently reported on an improved determination of the proton magnetic moment with a fractional precision of 300 parts in a trillion \cite{schneider2017science} and the first high-precision determination of the antiproton magnetic moment with a fractional precision of 1.5 parts in a billion \cite{smorra2017nature}. This measurement, based on a newly invented multi-trap method, improves the fractional precision achieved in previous studies \cite{Jack2012Proton,Jack2013Antiproton} by more than a factor of 3000.  These multi-trap based high-precision magnetic moment measurements on protons and antiprotons require low-noise conditions much more demanding than in any other ion trap experiment. Compared to experiments on electrons and positrons \cite{hanneke2008,vandyck1987}, the 660-fold smaller proton/antiproton magnetic moment makes it much more challenging to apply high-fidelity single particle spin-quantum spectroscopy techniques \cite{smorra2017}. Our experiments become possible only in cryogenic ultra-low-noise Penning-trap instruments, which provide energy stabilities of the particle motion on the peV/s range, effectively corresponding to a parasitic transition rate acceptance limit of, at most, two motional quanta over several minutes of measurement time.\\
In this Letter we report on the characterization of the electric field fluctuations in a cryogenic Penning trap by explicit measurements of cyclotron quantum transition rates of a single antiproton using the continuous Stern-Gerlach effect \cite{dehmelt1973}.
\begin{figure*}
	\centering
	\includegraphics[width=1\textwidth]{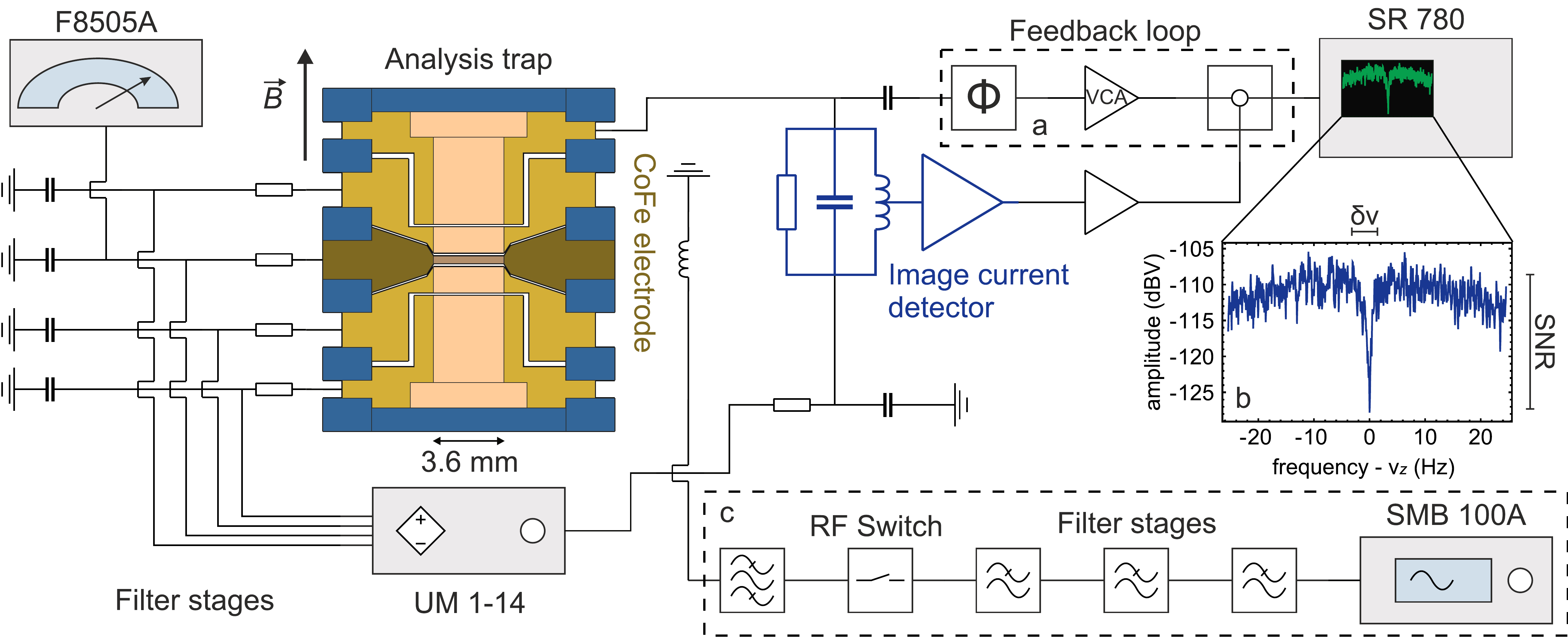}
	\caption{Experimental setup. The particle is confined inside a set of five cylindrical electrodes (golden/brown) with an inner diameter of 3.6\,mm. The central ring electrode is made out of a cobalt-iron alloy generating the magnetic inhomogeneity used for application of the continuous Stern-Gerlach effect \cite{dehmelt1973}. An ultra-stable voltage source (UM 1-14 by Stahl electronics) is connected to the trap electrodes via multi-stage low-pass filters. The central electrode voltage is simultaneously recorded by a FLUKE F8505A reference voltmeter. For axial frequency measurements, a feedback-cooled (a) image current detection system (blue) is used connected to an outer electrode \cite{nagahama2016}. The particles' axial oscillation frequency is obtained from the Fast Fourier Transformed detector spectrum (b). A Rohde\&Schwarz SMB 100A frequency generator equipped with high order low-pass and band-pass filters is used for particle manipulation (c).} 
	\label{fig:setup}
\end{figure*}
The observed electric field spectral noise density is more than two orders of magnitude lower than in room temperature Penning traps \cite{goodwin2016} and more than 1000 times smaller than observed in cryogenic Paul trap experiments \cite{brownnutt2015}. Based on heating rate measurements at various particle orbits we identify fluctuations in the trapping field caused by residual voltage noise as the dominant heating mechanism. Anomalous heating is not observed within our measurement resolution.\\
The measurements are conducted in the cryogenic spin-state analysis trap of the BASE apparatus at CERN \cite{smorra2015epj}, which is shown in Fig$.$ \ref{fig:setup}.
The Penning trap is realized using a superconducting magnet at $1.945$\,T combined with a quadrupolar electrostatic potential provided from a set of five carefully designed cylindrical electrodes with an inner diameter of 3.6\,mm \cite{CCRodegheri2012}. The central ring electrode is made out of a Co/Fe alloy, which distorts the nearly homogeneous axial magnetic field to ${B_z = B_0 + B_2 \left(z^2 - \rho^2/2 \right)}$, deliberately generating a magnetic inhomogeneity of $B_2~=~272\,\text{kT}\text{m}^{-1}$. The trajectory of a single antiproton stored in a Penning trap is composed of three harmonic oscillator modes. The modified cyclotron motion at $\nu_+$ and the magnetron motion at $\nu_-$ are perpendicular to the magnetic field, while the particle oscillates along the magnetic field lines with axial frequency $\nu_z$. For the BASE analysis trap, $\nu_+\approx 17.845\,$MHz, $\nu_-\approx10\,$kHz and $\nu_z\approx 675\,$kHz.\\
The gold-plated OFHC electrodes are placed inside an indium-sealed vacuum chamber which is cooled to ${T \approx 6\,}$K. Cryo-pumping provides an ultra-high vacuum with pressures $< 3 \times 10^{-18}$\,mbar which enables storage times $> 10\,a$ \cite{sellner2017}. Radio frequency (rf) lines equipped with high order low-pass and band-pass filters as well as high-insulation switches are used for particle manipulation (Fig$.\,$\ref{fig:setup}\,c). The axial oscillation frequency $\nu_{z}$ is measured by an  image current detection system \cite{nagahama2016}. 
The detector's time transient is processed with a Fast Fourier Transform (FFT) spectrum analyzer. Once cooled to thermal equilibrium, the particle signature appears as a dip in the resulting frequency spectrum \cite{wineland1975} (see Fig$.\,$\ref{fig:setup}\,b). A least-squares fit of the recorded spectra yields the axial frequency $\nu_z$. In the measurements reported here, we apply active electronic feedback cooling (see Fig$.$\,\ref{fig:setup}\,a) \cite{dehmelt1986,durso2003}, which enables measurements at low axial temperature ($T_z \approx 1{.}92(10)$\,K) and high axial frequency stability \cite{smorra2017}.\\
For explicit measurements of modified cyclotron transition rates we utilize the continuous Stern-Gerlach effect \cite{dehmelt1973}. Here, the interaction of the particle's magnetic moment $\mu_z=\mu_++\mu_-+\mu_s$  with the strong magnetic inhomogeneity $B_2$ results in a magnetostatic axial energy $E_{B,z}=-\mu_z\times {B}_z(z)$, where $\mu_+$ and $\mu_-$ are the angular magnetic moments associated with the modified cyclotron and the magnetron mode, while $\mu_s$ is the spin magnetic moment. As a result, the antiproton's axial frequency $\nu_{z}=\nu_{z,0}+\Delta\nu_{z}$ becomes a function of the radial quantum states:
\begin{equation}
\begin{split}
&\Delta \nu_z(n_+,n_-,m_s)\\&=\frac{h \nu_+}{4\pi^2 m_{\bar{p}}\nu_{z}}\frac{B_2}{B_0} \left[\left(n_++\frac{1}{2}\right)+\frac{\nu_-}{\nu_+}\left(n_-+\frac{1}{2}\right)+\frac{g_{\bar{p}} m_s}{2}\right].
\end{split}
\label{axialfreqshiftformula}
\end{equation}
Transitions in the corresponding states $\left(m_s, n_+, n_-\right)$ lead to axial frequency shifts of {$\Delta \nu_{z,\text{s}}~=~172(10)\,$}mHz, ${\Delta \nu_{z,\text{+}}~=~}$62(4)\,mHz and ${\Delta \nu_{z,-}~=~}$40(3)\,$\mu$Hz, respectively.\\
To determine the transition rate $\zeta_+$ of the cyclotron motion we first prepare a particle at low radial energy with $n_+<200$ \cite{mooser2013}. Then, we record sequences of axial frequency measurements $\nu_{z,k}$ with an averaging time $\tau_0~=~50$\,s. Subsequently, we evaluate the standard deviation $\sigma_{\nu_z}(\tau)=\sigma \left(\langle \nu_{z,j+1} \rangle \left(\tau\right)- \langle \nu_{z,j} \rangle \left(\tau\right)\right)$, where $\langle \nu_{z,j} \rangle(\tau)$ represents the mean values of a sub-series of axial frequency measurements with an averaging time $\tau = l \times \tau_0$. A result of such an overlapping differential Allan deviation $\sigma_{\nu_z}(\tau)$ \cite{allan1966} is shown in Fig$.$ \ref{fig:complete_adevplot_20161130_2325_p1} as blue filled circles.
Various measured and simulated contributions to $\sigma_{\nu_z}(\tau)$ are also plotted in Fig. 2. The contribution from voltage fluctuations $\sigma_v(\tau)$ (dark red triangles) is extracted from simultaneous measurement of the voltage supply stability, as shown in Fig 1. The contribution from white frequency measurement noise, $\sigma_{\text{FFT}}(\tau) \propto \delta \nu_z^{1/2} $\,$\text{SNR}^{-1/4}$ (dark red squares) is calculated \cite{smorra2015epj}, $\delta \nu_z$ being the line-width of the axial frequency dip and SNR the signal-to-noise ratio (see Fig$.$\,\ref{fig:setup} b). At small averaging times ($\tau<100\,$s), these two contributions dominate.
Meanwhile, with long averaging times ($\tau>250\,$s), $\sigma_{\nu_z}(\tau)$ is dominated by transition rates $\zeta_+$ in the modified cyclotron mode,
\begin{equation}
\sigma_{\nu_z}(\tau)\propto\sqrt{\sigma_v(\tau)^2+\sigma_{\text{FFT}}(\tau)^2+\tau \left(\Delta\nu^2_{z,+} \zeta_+\right)}.
\label{adevporportionalities}
\end{equation} 
By analyzing such data and comparing the Allan deviation to Monte-Carlo simulated noise-driven random walks, we extract an absolute cyclotron transition rate of $\zeta_+ = 6(1)\,\text{h}^{-1}$, see Fig$.$~\ref{fig:complete_adevplot_20161130_2325_p1}. Note that $\zeta_+$ describes a nearly undirected random walk.
\begin{figure}
	\centering
	\includegraphics[width=1\linewidth]{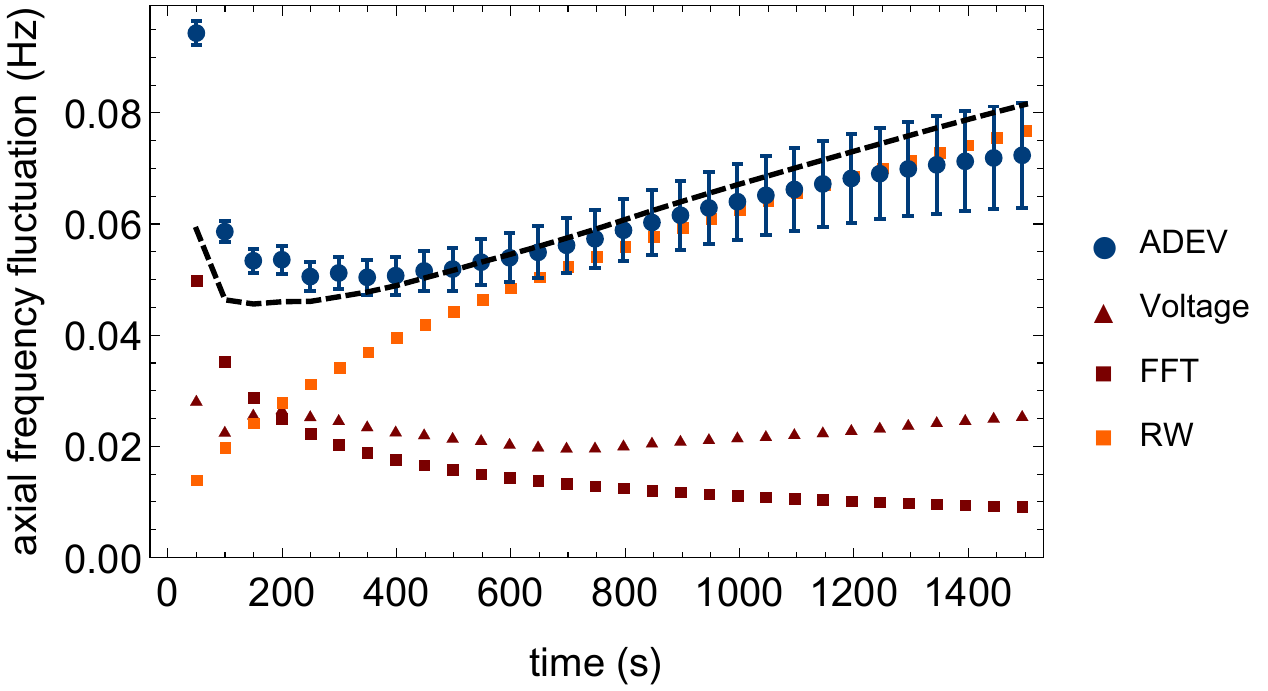}
	\caption{Axial frequency stability analysis for an antiproton at low radial energy. The differential Allan deviation  $\sigma_{\nu_z}\left(\tau \right)$ is displayed in blue. Frequency measurement noise (FFT, dark red squares) and voltage fluctuations (dark red triangles) contribute to the observed frequency instability. Contributions from a simulated random walk of the cyclotron energy are displayed in orange. The dashed black line is given by the sum of random walk, FFT and voltage contributions. For time spans larger than 250\,s, the Allan deviation is dominated by random walks, $\zeta_+ = 6(1)\,\text{h}^{-1}$. Transition rate uncertainties are extracted from the $\sigma_{\nu_z}\left(\tau \right)$-error bars. For this dataset, 900 frequency measurements were conducted over 12\,h.}
	\label{fig:complete_adevplot_20161130_2325_p1}
\end{figure}
The observed transition rates can be related to the noise spectral density of the radial electric field $S_E(\omega_+)$ at the modified cyclotron frequency. Considering first order transitions in a noise-driven quantum mechanical oscillator \cite{savard1997}, cyclotron transitions rates are given by  
\begin{equation}
\zeta_+=\frac{q^2 n_{+}}{2 m_{\bar{p}}\hbar \omega_{+}} S_E\left(\omega_{+}\right),
\label{transitionrateformula}
\end{equation}
where $S_{E}(\omega_{+})$ is the spectral density of electric field noise acting on the particle's cyclotron motion. The average increase of $n_+$ is given by the heating rate d$\overline{n}_+/\text{d}t~=~\zeta_+~\times~1/(2n_+)$ for $n_+ \gg 1$. 
Together with the determination of a lower limit for $n_+$ based on the continuous Stern-Gerlach effect \cite{ulmer2011prlobservation} we obtain an upper limit for the electric field spectral density of $S_{E}(\omega_+) \leq7.5^{+3.4}_{-2.8}\times 10^{-20}\,\text{V}^2\text{m}^{-2}\text{Hz}^{-1}$.
The absolute resolution of our axial frequency measurements is limited by environmental variations of temperature, cryoliquid levels, and pressure, which impose uncertainties on the determination of both the cyclotron quantum number $n_+$ as well as the transition rate $\zeta_+$.
\begin{figure}
	\centering
	\includegraphics[width=1\linewidth]{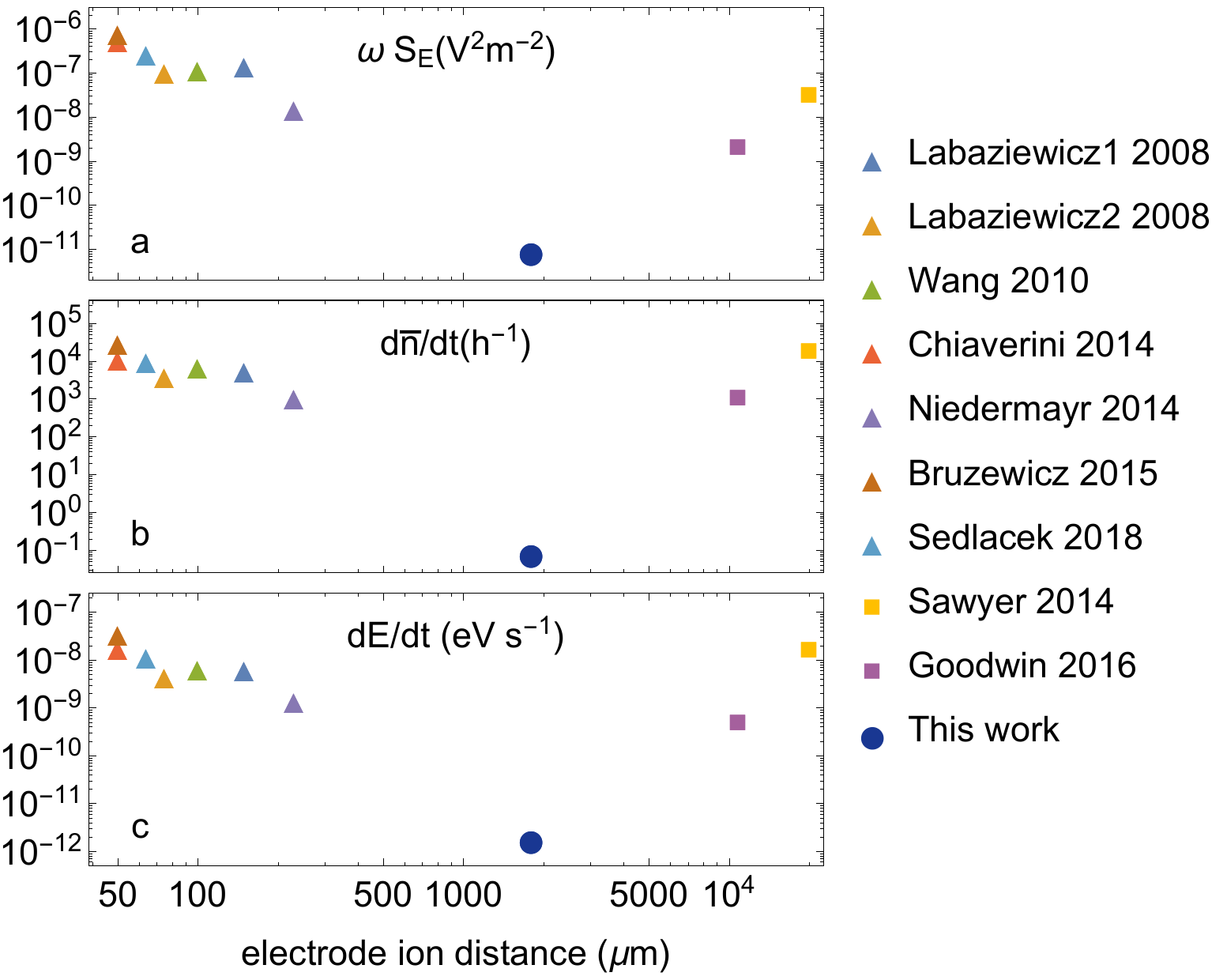}
	\caption{Single particle stabilities as a function of the electrode-to-ion distance $d$. Fig$.$ a) displays the electric-field noise spectral density $S_E(\omega)$ scaled by angular trap frequency $\omega$, Fig$.$ b) depicts heating rates d$\overline{n}/$d$t$ and in Fig$.$ c) the energy increase $\text{d}E/\text{d}t$  is shown. The triangles represent measurements performed in cryogenic 2D-Paul traps \cite{labaziewicz2008_1,labaziewicz2008_2,wang2010,chiaverini2014,niedermayr2014,bruzewicz2015,sedlacek2018}, squares denote measurements in Penning traps on single ions \cite{goodwin2016} and ion crystals \cite{sawyer2014,stutter2018} conducted at room temperature. This work is plotted as a blue circle.}
	\label{fig:compplot}
\end{figure}
Nevertheless, our upper limit for $S_{E}(\omega_+)$ is far below the results reported by cryogenic Paul trap  \cite{labaziewicz2008_1,labaziewicz2008_2,wang2010,chiaverini2014,niedermayr2014,bruzewicz2015,sedlacek2018} and room temperature Penning-trap experiments \cite{goodwin2016,sawyer2014,stutter2018}. The current best limits extracted from those experiments are $S_{E}(\omega) = 2.4 \times 10^{-15}$\,$\text{V}^2\text{m}^{-2}\text{Hz}^{-1}$ \cite{niedermayr2014,brownnutt2015} and $S_{E}(\omega) = 8 \times 10^{-16}\,\text{V}^2\text{m}^{-2}\text{Hz}^{-1}$ \cite{goodwin2016,stutter2018}, respectively. Fig$.$~\ref{fig:compplot}~a) displays the commonly used scaled electric field noise $\omega S_E\left(\omega\right)$ which accounts for the $1/\omega-$dependence of the heating rate \cite{turchette2000,brownnutt2015}. Our result $\omega S_E\left(\omega\right) \leq 8.8^{+4.0}_{-3.2} \times 10^{-12}\,\text{V}^2\text{m}^{-2}$ sets an upper limit which is a factor of 1800 \cite{niedermayr2014} lower than the best reported Paul trap heating rates and a factor of 230 lower than the best Penning trap
\cite{goodwin2016}. Fig$.$~\ref{fig:compplot}~b) plots the heating rate $\text{d}\overline{n}/\text{d}t$ for various experiments, which is in our case below $0.1\,\text{h}^{-1}$. The corresponding energy increase $\text{d}E/\text{d}t$, plotted in Fig$.$~\ref{fig:compplot}~c), is on the order of peV/s,  demonstrating to our knowledge the highest energy stability of a particle in any ion trap experiment.\\
To further investigate the residual drive mechanism, we measure transition rates $\zeta_+(\rho_-)$ as a function of the particle's magnetron radius $\rho_-$, thereby changing the trapping field at the particle position. We excite the magnetron mode and record series of axial frequency sequences $\Omega_k(\nu_z,\rho_-)$ for in total 7 different magnetron radii, thereby tracing a radial range of $6\,\mu\text{m} \leq \rho_- \leq 65\,\mu$m.
\begin{figure}
	\centering
	\includegraphics[width=1\linewidth]{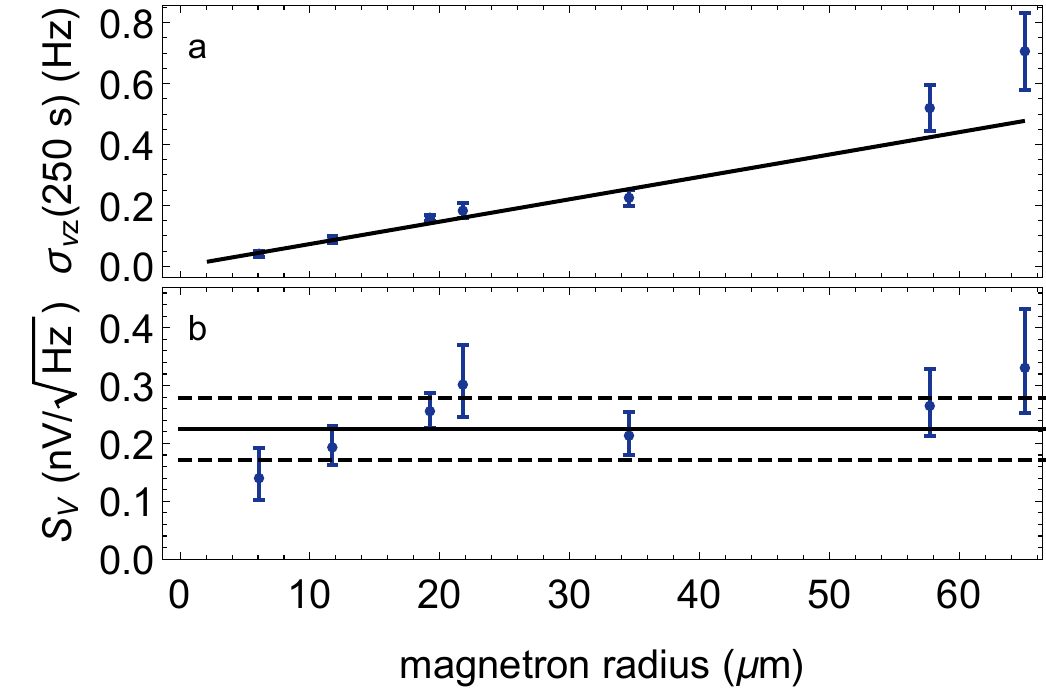}
	\caption{Results of frequency stability measurements for particles at different magnetron radii $\rho_-$. (a): Measured Allan deviation $\sigma_{\nu_z}\left(\tau\right)$ of the axial frequency for an averaging time of $\tau = 250\,\text{s}$. The black line denotes calculated values for $\sigma_{\nu_z}\left(\tau\right)$ assuming transition rates are linked to trap voltage fluctuations. (b): Calculated electrode voltage fluctuations $S_V$. The linear increase of $\sigma_{\nu_z}\left(\tau\right)$ is in good agreement with cyclotron transition rates driven by trapping voltage fluctuations. The extracted voltage fluctuation $S_V\left(\omega_+\right)$ (black lines) are constant for $6\,\mu\text{m}\leq \rho_- \leq 65\,\mu\text{m}$, confirming that they can be regarded as the dominant source of electric field fluctuations in the trap.
}
	\label{fig:adevvoltnoisevsmagrad}
\end{figure}
The results of these measurements are displayed in Fig$.$ \ref{fig:adevvoltnoisevsmagrad}. In Fig$.$ \ref{fig:adevvoltnoisevsmagrad}~a) we show the measured axial frequency fluctuation $\sigma_{\nu_z}\left(\rho_-,\tau=250\,\text{s}\right)$. For the data points displayed in Fig$.$ \ref{fig:adevvoltnoisevsmagrad}~b), we analyze the transition rate $\zeta_+(\rho_-)$ of each dataset $\Omega_k(\nu_z,\rho_-)$ and determine the spectral density $S_{V}\left(\omega_{+}\right)$ of an equivalent effective voltage noise source present on each trap electrode:
\begin{equation}
S_E\left(\omega_{+}\right)=\Lambda^2\left(\rho,z\right) S_V\left(\omega_{+}\right),
\label{voltagenoiseformula}
\end{equation}
where $\Lambda \left(\rho,z\right)$ describes the relation between the electric field at the particle position $\vec{x}=\left(\rho,z\right)$ and the potential $V_n$ created by the $n$-th electrode:
\begin{equation}
\Lambda^2\left(\rho,z\right)=\sum^{5}_{n=1}\left(\frac{\partial V_n}{\partial \rho}\right)^2\propto \rho^2,
\label{lambdaformula}
\end{equation}
for low cyclotron energies, $\rho \approx \rho_-$.
The linear increase of $\sigma_{\nu_z}\left(\tau\right)\propto \rho_-$ observed in Fig$.$~\ref{fig:adevvoltnoisevsmagrad}~a) reflects a quadratic increase of transition rates $\zeta_+\propto \rho_-^2$ (Eq$.$\,\ref{adevporportionalities}). This is expected from Eq$.$\,(\ref{transitionrateformula}, \ref{voltagenoiseformula}, \ref{lambdaformula}), assuming electrode voltage noise $S_V$ as the dominant source of electric field fluctuations. We obtain $S_V = 225(54)\,\text{pV}\,\text{Hz}^{-1/2}$. Anomalous heating reported from Paul traps \cite{turchette2000,brownnutt2015} scales with $d^{-4}$, $d$ denoting the electrode-ion-distance. Since the variation of $d$ is small ($\Delta d/d = 1/60$) for the considered magnetron radii, anomalous heating would result in a nearly constant electric field noise spectral density. Since a clear increase is observed in $\zeta_+$, anomalous heating is ruled out as the dominant heating mechanism. Its effect is constrained to be below $S_E\left(\omega_+\right) \leq 7.5(3.4)\times 10^{-20}\,\text{V}^{2}\text{m}^{-2} \text{Hz}^{-1}$.\\
\begin{table}
\centering
\vspace{5mm}
\begin{tabular}{|c|c|}
	\hline 
	Observed $S_V$ & $ 225(54)\,\text{pV}\,\text{Hz}^{-1/2}$ \\ 
    \hline \hline
	Axial detection system & $ 1.5\,\text{pV}\,\text{Hz}^{-1/2}$ \\ 
	\hline 
	RC filter stages & $< 1\,\text{pV}\,\text{Hz}^{-1/2}$ \\ 
	\hline 
	Electrode Johnson noise & $\sim 3 \times 10^{-3}\,\text{pV}\,\text{Hz}^{-1/2}$ \\ 
	\hline \hline
	Blackbody radiation & $\omega_+ \times S_{E}\left(\omega_+\right) \sim 6 \times 10^{-14}\,\text{V}^2\text{m}^{-2}$ \\ 
	\hline 
	Background pressure & $\zeta_+ < 4 \times 10^{-9}\,\text{s}^{-1}$ \\ 
	\hline 
\end{tabular}
\caption{Parasitic voltage fluctuation and heating rate contributions.}
\end{table} 
In order to investigate contributions to $S_V$ we consider the experimental setup depicted in Fig$.$\,\ref{fig:setup}. The effective parallel resistance of the axial detection system at the cyclotron frequency contributes about $1.5\,\text{pV}\,\text{Hz}^{-1/2}$. The Johnson noise of the electrode RC-filters is below $1\,\text{pV}\,\text{Hz}^{-1/2}$, the electrode Johnson noise is on the order of $10^{-3}\,\text{pV}\,\text{Hz}^{-1/2}$. None of these mechanisms can explain the observed voltage fluctuations.
Field fluctuations arising from blackbody radiation are estimated to be $\omega_+ \times S_E^{(\text{BB})} \approx 6 \times 10^{-14}\,\text{V}^2\text{m}^{-2}$ \cite{henkel1999,brownnutt2015}, which is two orders of magnitude lower than our limit of $\omega S_E\left(\omega\right) \leq 8.8^{+4.0}_{-3.2} \times 10^{-12}\,\text{V}^2\text{m}^{-2}$. A trapped ion polarizes neutral background gas atoms and thereby induces collisions described by the Langevin rate $\gamma$, which is proportional to the background gas density \cite{wineland1998,chiaverini2014}. From our antiproton lifetime measurement \cite{sellner2017} we derived upper limits for the partial pressure of hydrogen $p_{\text{upper},H} < 1.2 \times 10^{-18}\,$mbar and helium $p_{\text{upper},He} < 2.7 \times 10^{-18}$\,mbar leading to $\zeta_+ < 4 \times 10^{-9}\text{s}^{-1}$. Voltage supply (UM1-14) noise at $\nu_+$ is ruled out by independent measurements.
Therefore we assume parasitic coupling of stray EMI noise onto the trap electrodes to be the dominant source of electric field fluctuations in our trap. A further improvement to achieve even lower heating rates which will enhance the sensitivity of our experiment will be subject of future experimental studies.\\
We acknowledge financial support of RIKEN Pioneering Project Funding, RIKEN FPR program, RIKEN JRA program, the Max-Planck Society, the CERN fellowship program, the EU (Marie Sklodowska-Curie grant agreement No. 721559) and the KAS/BMBF PhD fellowship program. We acknowledge support from CERN, in particular from the AD operation team.


\begin{thebibliography}{}
\bibitem{schmidt2005} P. O. Schmidt, T. Rosenband, C. Langer, W. M. Itano, J. C. Bergquist, and D. J. Wineland, Science \textbf{309}, 749 (2005).
\bibitem{kitching2011} J. Kitching, S. Knappe and E. A. Donley,  IEEE Sens. J. \textbf{11}, 1749 (2011).
\bibitem{blatt2008} R. Blatt, and D. J. Wineland, Nature \textbf{453}, 1008 (2008).
\bibitem{turchette2000} Q. A. Turchette, D. Kielpinski, B. E. King, D. Leibfried, D. M. Meekhof, C. J. Myatt, M. A. Rowe, C. A. Sackett, C. S. Wood, W. M. Itano, C. Monroe, and D. J. Wineland, Phys. Rev. A \textbf{61}, 063418 (2000).
\bibitem{brownnutt2015} M. Brownnutt, M. Kumph, P. Rabl, and R. Blatt \textit{et al.}, Rev. Mod. Phys. \textbf{87}, 1419 (2015).
\bibitem{kosteleckyPenning} R. Bluhm, V. A. Kosteleck\'y, and N. Russell, Phys. Rev. D \textbf{57}, 3932 (1998).
\bibitem{safronova2018} M. S. Safronova, D. Budker, D. DeMille, D. F. J. Kimball, A. Derevianko, and C. W. Clark, Rev. Mod. Phys. \textbf{90}, 025008 (2018).
\bibitem{sturm2014} S. Sturm, F. K\"ohler, J. Zatorski, A. Wagner, Z. Harman, G. Werth, W. Quint, C. H. Keitel, and K. Blaum, Nature \textbf{506}, 467 (2014).
\bibitem{sturm2011} S. Sturm, A. Wagner, B. Schabinger, and K. Blaum, Phys. Rev. Lett. \textbf{107}, 143003 (2011).
\bibitem{hanneke2008} D. Hanneke, S. Fogwell, and G. Gabrielse, Phys. Rev. Lett. \textbf{100}, 120801 (2008).
\bibitem{vandyck1987} R. S. Van Dyck, P. B. Schwinberg, and H. G. Dehmelt, Phys. Rev. Lett. \textbf{59}, 26 (1987).
\bibitem{cpt} G. L\"uders, Ann. Phys. \textbf{2}, 1 (1957).
\bibitem{dehmelt1999} H. Dehmelt, R. Mittleman, R. S. Van Dyck, and P. Schwinberg, Phys. Rev. Lett. \textbf{83}, 4694 (1999).
\bibitem{smorra2015epj} C. Smorra \textit{et al.}, Eur. Phys. J. Special Topics \textbf{224}, 3055 (2015).
\bibitem{smorra2017nature} C. Smorra \textit{et al.}, Nature \textbf{550}, 371 (2017).
\bibitem{schneider2017science} G. Schneider \textit{et al.}, Science \textbf{358}, 1081 (2017).
\bibitem{Jack2012Proton} J. DiSciacca and G. Gabrielse, Phys. Rev. Lett. \textbf{108}, 153001 (2012).
\bibitem{Jack2013Antiproton} J. DiSciacca, M. Marshall, K. Marable, G. Gabrielse, S. Ettenauer, E. Tardiff, R. Kalra, D. W. Fitzakerley, M. C. George, E. A. Hessels, C. H. Storry, M. Weel, D. Grzonka, W. Oelert, and T. Sefzick, Phys. Rev. Lett. \textbf{110}, 130801 (2013).
\bibitem{smorra2017} C. Smorra \textit{et al.}, Phys. Lett. B \textbf{769}, 1 (2017).
\bibitem{dehmelt1973} H. Dehmelt and P. Ekstr\"om, Bull. Am. Phys. Soc. \textbf{18}, 727 (1973).
\bibitem{goodwin2016} J. F. Goodwin, G. Stutter, R. C. Thompson, and D. M. Segal, Phys. Rev. Lett. \textbf{116}, 143002 (2016).
\bibitem{CCRodegheri2012} C. C. Rodegheri, K. Blaum, H. Kracke, S. Kreim, A. Mooser, W. Quint, S. Ulmer, and J. Walz, New J. Phys. \textbf{14}, 063011 (2012).
\bibitem{sellner2017} S. Sellner \textit{et al.}, New J. Phys. \textbf{19}, 083023 (2017).
\bibitem{nagahama2016} H. Nagahama \emph{et al.}, Rev. Sci. Instrum. \textbf{87}, 113305 (2016).
\bibitem{wineland1975} D. J. Wineland and H. G. Dehmelt, J. Appl. Phys. \textbf{46}, 919 (1975).
\bibitem{dehmelt1986} H. Dehmelt, W. Nagourney, and J. Sandberg, Proc. Natl. Acad. Sci. USA \textbf{83}, 5761 (1986).
\bibitem{durso2003} B. D'Urso, B. Odom, and G. Gabrielse, Phys. Rev. Lett. \textbf{90}, 043001 (2003).
\bibitem{mooser2013} A. Mooser, H. Kracke, K. Blaum, S. A. Br\"auninger, K. Franke, C. Leiteritz, W. Quint, C. C. Rodegheri, S. Ulmer, and J. Walz, Phys. Rev. Lett. \textbf{110}, 140405 (2013).
\bibitem{allan1966} D. W. Allan, Proc. of the IEEE \textbf{54}, 221 (1966).
\bibitem{savard1997} T. A. Savard, K. M. O'Hara, and J. E. Thomas, Phys. Rev. A \textbf{56}, R1095 (1997).
\bibitem{ulmer2011prlobservation} S. Ulmer, C. C. Rodegheri, K. Blaum, H. Kracke, A. Mooser, W. Quint, and J. Walz, Phys. Rev. Lett. \textbf{106}, 253001 (2011).
\bibitem{labaziewicz2008_1} J. Labaziewicz, Y. Ge, P. Antohi, D. Leibrandt, K. R. Brown, and I. L. Chuang, Phys. Rev. Lett. \textbf{100}, 013001 (2008).
\bibitem{labaziewicz2008_2} J. Labaziewicz, Y. Ge, D. R. Leibrandt, S. X. Wang, R. Shewmon, and I. L. Chuang, Phys. Rev. Lett. \textbf{101}, 180602 (2008).
\bibitem{wang2010} S. X. Wang, Y. Ge, J. Labaziewicz, E. Dauler, K. Berggren, and I. L. Chuang, Appl. Phys. Lett. \textbf{97}, 244102 (2010).
\bibitem{chiaverini2014} J. Chiaverini and J. M. Sage, Phys. Rev. A \textbf{89}, 012318 (2014).
\bibitem{niedermayr2014} M. Niedermayr, K. Lakhmanskiy, M. Kumph, S. Partel, J. Edlinger, M. Brownnutt, and R. Blatt, New J. Phys. \textbf{16}, 113068 (2014).
\bibitem{bruzewicz2015} C. D. Bruzewicz, J. M. Sage, and J. Chiaverini, Phys. Rev. A \textbf{91}, 041402(R) (2015).
\bibitem{sedlacek2018} J. A. Sedlacek, A. Greene, J. Stuart, R. McConnell, C. D. Bruzewicz, J. M. Sage, and J. Chiaverini, Phys. Rev. A \textbf{97}, 020302(R) (2018).
\bibitem{sawyer2014} B. C. Sawyer, J. W. Britton, and J. J. Bollinger, Phys. Rev. A \textbf{89}, 033408 (2014).
\bibitem{stutter2018} G. Stutter, P. Hrmo, V. Jarlaud, M. K. Joshi, J. F. Goodwin, and R. C. Thompson, J. Mod. Opt. \textbf{65}, 540 (2018).
\bibitem{henkel1999} C. Henkel, S. P\"otting, and M. Wilkens, Appl. Phys. B \textbf{69}, 379 (1999).
\bibitem{wineland1998} D. J. Wineland, C. Monroe, W. M. Itano, D. Leibfried, B. E. King, and D. M. Meekhof, J. Res. Natl. Inst. Stand. Technol. \textbf{103}, 259 (1998).
\end{thebibliography}
\end{document}